\documentclass[useAMS,usenatbib,usegraphicx]{mn2e}
\usepackage{graphicx}
\usepackage[breaklinks,colorlinks,citecolor=black,linkcolor=black,urlcolor=black]{hyperref}

\newcommand*{\mysub}[2]{\ensuremath{#1_{\mathrm{#2}}}}
\newcommand*{\unit}[1]{\ensuremath{\mathrm{\, #1}}}

\newcommand*{\Msun}{\ensuremath{\, M_{\odot}}}

\newcommand*{\keV}{\unit{keV}}
\newcommand*{\erg}{\unit{erg}}

\newcommand*{\km}{\unit{km}}
\newcommand*{\Mpc}{\unit{Mpc}}
\newcommand*{\second}{\unit{s}}

\newcommand*{\Mgas}{\mysub{M}{gas}}

\newcommand*{\Lce}{\mysub{L}{ce}}
\newcommand*{\Lm}{\mysub{L}{M07}}
\newcommand*{\Yx}{\mysub{Y}{X}}

\newcommand*{\LCDM}{\ensuremath{\Lambda}CDM}
\newcommand*{\Omegam}{\mysub{\Omega}{m}}

\newcommand*{\rhocr}{\mysub{\rho}{cr}}

\newcommand*{\E}[1]{\ensuremath{\times 10^{#1}}}

\newcommand*{\ltsim}{\ {\raise-.75ex\hbox{$\buildrel<\over\sim$}}\ }
\newcommand*{\gtsim}{\ {\raise-.75ex\hbox{$\buildrel>\over\sim$}}\ }
\newcommand*{\proptosim}{\ {\raise-.75ex\hbox{$\buildrel\propto\over\sim$}}\ }

\newcommand*{\Chandra}{{\it Chandra}}
\newcommand*{\Athena}{{\it Athena}}
\newcommand*{\Lynx}{{\it Lynx}}

\defcitealias{Mantz1606.03407}{M16}
\newcommand*{\pfive}{\citetalias{Mantz1606.03407}}

\newcommand*{\figscale}{0.9}

\begin{document}

\title[Center-Excised Luminosity as a Cluster Mass Proxy]{Center-Excised X-ray Luminosity as an Efficient Mass Proxy for Future Galaxy Cluster Surveys}

\author[A. B. Mantz et al.]{Adam B. Mantz,$^{1,2}$\thanks{E-mail: \href{mailto:amantz@slac.stanford.edu}{\tt amantz@slac.stanford.edu}} {}
  Steven W. Allen,$^{1,2,3}$
  R. Glenn Morris,$^{1,3}$
  Anja von der Linden$^4$
  \\$^1$Kavli Institute for Particle Astrophysics and Cosmology, Stanford University, 452 Lomita Mall, Stanford, CA 94305, USA
  \\$^2$Department of Physics, Stanford University, 382 Via Pueblo Mall, Stanford, CA 94305, USA
  \\$^3$SLAC National Accelerator Laboratory, 2575 Sand Hill Road, Menlo Park, CA  94025, USA
  \\$^4$Department of Physics and Astronomy, Stony Brook University, Stony Brook, NY 11794, USA
}
\date{Submitted 25 May 2017. Accepted 26 September 2017.}

\pagerange{\pageref{firstpage}--\pageref{lastpage}} \pubyear{2017}
\maketitle
\label{firstpage}

\begin{abstract}
  The cosmological constraining power of modern galaxy cluster catalogs can be improved by obtaining low-scatter mass proxy measurements for even a small fraction of sources. In the context of large upcoming surveys that will reveal the cluster population down to the group scale and out to high redshifts, efficient strategies for obtaining such mass proxies will be valuable. In this work, we use high-quality weak lensing and X-ray mass estimates for massive clusters in current X-ray selected catalogs to revisit the scaling relations of the projected, center-excised X-ray luminosity (\Lce{}), which previous work suggests correlates tightly with total mass. Our data confirm that this is the case, with \Lce{} having an intrinsic scatter at fixed mass comparable to that of gas mass, temperature or $\Yx$. Compared to these other proxies, however, \Lce{} is less susceptible to systematic uncertainties due to background modeling, and can be measured precisely with shorter exposures. This opens up the possibility of using \Lce{} to estimate masses for large numbers of clusters discovered by new X-ray surveys (e.g.\ eROSITA) directly from the survey data, as well as for clusters discovered at other wavelengths with relatively short follow-up observations. We describe a simple procedure for making such estimates from X-ray surface brightness data, and comment on the spatial resolution required to apply this method as a function of cluster mass and redshift. We also explore the potential impact of \Chandra{} and XMM-{\it Newton} follow-up observations over the next decade on dark energy constraints from new cluster surveys.
\end{abstract}


\section{Introduction} \label{sec:intro}

Measurements of observable quantities that correlate tightly with total mass, i.e.\ that function as effective low-scatter proxies, can significantly boost the cosmological constraining power of galaxy cluster surveys (\citealt{Wu0907.2690}; \citealt*{Allen1103.4829}). Key considerations are the intrinsic scatter in the mass proxy at fixed true mass and the observing resources required to measure the proxy.

Among X-ray mass proxies, total X-ray luminosity is the most straightforward to measure, requiring (in addition to the cluster redshift) only tens of source counts, and comes ``for free'' from surveys which identify clusters through their X-ray emission. Other proxies such as gas mass and temperature require hundreds to thousands of X-ray source counts, but have a smaller intrinsic scatter than total luminosity. While the relatively large ($\sim40$ per cent) scatter in total X-ray luminosity at fixed mass limits its utility as a mass proxy, it has been recognized that X-ray luminosity measured in an annulus, excluding the cluster center, has a significantly smaller intrinsic scatter, while retaining the attractive simplicity of a luminosity measurement (e.g.\ \citealt{Maughan0703504, Zhang0702739, Pratt0809.3784, Mantz0909.3099, Mantz1509.01322}).

Physically, the reason why excising cluster centers reduces the intrinsic scatter is easy to understand. The intracluster medium shows great variation in cluster centers, primarily driven by the development of bright, dense ``cool cores'' of gas, associated with sharp density and surface brightness peaks and reduced temperatures, in a fraction of the cluster population (e.g.\ \citealt{Fabian1994MNRAS.267..779F, Markevitch9802059}). In contrast, the gas density and temperature profiles outside of cluster centers are remarkably similar (e.g.\ \citealt{Vikhlinin0507092, Allen0706.0033, Croston0801.3430, Pratt0809.3784, Mantz1509.01322}).

For massive (hot) clusters, the soft X-ray bremsstrahlung emissivity, and the associated K corrections, are nearly independent of temperature. For concreteness, we define the ``soft'' X-ray band to be 0.1--2.4\,keV (rest frame), and ``hot'' temperatures to mean $\gtsim 4$\,keV; the important general features here are that bremsstrahlung continuum emission dominates over line emission for a ``hot'' source, and that the (redshifted) exponential cut-off of the bremsstrahlung spectrum falls outside of the ``soft'' energy band (see more detailed discussion in Section~\ref{sec:applicability}). In the appropriate regime, the soft-band surface brightness can thus be considered a relatively simple function of the gas density, with similarity in density profiles outside cluster centers translating into a small intrinsic scatter in both center-excised soft-band luminosity (hereafter \Lce{}) and integrated gas mass ($\Mgas$). In this context, a key difference between \Lce{} and \Mgas{} is that the former is a straightforward integral of the projected, K-corrected surface brightness in a given annulus, which is dominated by the region where the brightness is highest, usually the smallest radii in the annulus of integration. In contrast, the gas mass within a sphere is  a volume-weighted integral, and is thus dominated by the largest radii in the integration, where the signal-to-noise is typically lowest. Furthermore, estimating $\Mgas$ within a sphere requires knowledge of the gas profile at yet larger radii, so that projected emission can be accounted for.

While center-excised temperature measurements have become common for mass estimation (e.g.\ \citealt{Kravtsov0603205, Vikhlinin0805.2207}), the same cannot be said for \Lce{}. Yet, the existing evidence points to \Lce{} being potentially a useful, not to mention relatively cheap, mass proxy. To the extent that density profiles outside of cluster centers are self-similar, the intrinsic scatter in \Lce{} should be comparable to that of \Mgas{} and temperature ($\sim10$--15 per cent). For perfectly similar profiles measured with high signal-to-noise, it would provide identical information to \Mgas{}. In practice, using a signal that is most sensitive to relatively small radii (while still excluding the core) is an advantage, because the impact of the X-ray background and the likelihood of having limited azimuthal coverage of the cluster (at relatively low redshifts) both increase with radius. Similar comments apply to the comparison of \Lce{} and temperature, the latter being more sensitive to background modeling than either luminosity or gas mass. In addition, cross-calibration studies indicate good agreement in soft X-ray flux measurements between \Chandra{} and XMM-{\it Newton}, whereas temperature measurements for hot clusters with these two workhorse telescopes remain discrepant \citep{TsujimotoAA...525A..25, Schellenberger1404.7130}.

Motivated by these considerations, we revisit in this work the \Lce{}--mass scaling relation, using a relatively large sample of massive clusters for which we have high quality X-ray and/or weak gravitational lensing mass estimates. Section~\ref{sec:data} describes these data, while Section~\ref{sec:scaling} presents the scaling relations linking \Lce{} to gas mass and total mass, as probed by weak lensing. In Section~\ref{sec:discussion}, we outline the procedure for applying the $\Lce$--mass relation to obtain mass estimates of new clusters, address the applicability of \Lce{} (and center-excised measurements in general) in the context of the spatial resolution provided by current and planned X-ray observatories, and discuss the potential of \Lce{} observing programs to boost cosmological constraints from clusters. We conclude in Section~\ref{sec:conclusion}.

We define the characteristic radius and mass of a cluster in the conventional way, with respect to the critical density at its redshift,
\begin{equation} \label{eq:massdef}
  M_{\Delta} = \frac{4\pi}{3} \Delta \rhocr(z) r_\Delta^3,
\end{equation}
with $\Delta=500$. For brevity of notation, we will forgo subscripts ``500'' for mass, $\Mgas$, etc., but all quantities measured in this work are referenced to $r_{500}$. We assume a flat \LCDM{} cosmology with parameters $H_0=70\km\second^{-1}\Mpc^{-1}$ and $\Omegam=0.3$ throughout.

\section{Data} \label{sec:data}

Our data set consists of 139 clusters with \Chandra{} X-ray follow-up observations, originally selected from the ROSAT All-Sky Survey (RASS). The \Chandra{} data, occasionally supplemented by ROSAT PSPC surface brightness information, are used to extract integrated observables such as gas mass, average temperature, and total X-ray luminosity, which are known to correlate with total mass, with different amounts of intrinsic scatter. The analysis of these X-ray data, and their scaling relations with mass, is described in detail in \citet[][hereafter \pfive]{Mantz1606.03407}. Here we consider an additional mass proxy, \Lce{}. To be precise, this refers to the intrinsic (unabsorbed), rest-frame 0.1--2.4\,keV band\footnote{The exact choice of energy band is somewhat arbitrary. While the 0.1--2.4\,keV band provides continuity with luminosity measurements going back to the RASS, similar bands that have been used in the literature are in principle equally useful for our purposes. In our results (Table~\ref{tab:fits}), we also provide scaling relations appropriate for the commonly used 0.5--2.0\,keV band.} X-ray luminosity of a cluster, projected within an annular aperture with an inner radius of $0.15\,r_{500}$ and an outer radius of $r_{500}$. The particular choice of inner radius is intended to comfortably exclude the variability observed in cluster centers (e.g.\ \citealt{Mantz1509.01322}); we comment on the use of larger excluded regions in Section~\ref{sec:psf}.

In addition, we use mass estimates from weak gravitational lensing data for 50 clusters from the Weighing the Giants project. The analysis of the lensing data, extraction of mass estimates, and accounting for systematic uncertainties are discussed by \citet{von-der-Linden1208.0597}, \citet{Kelly1208.0602} and \citet{Applegate1208.0605}.

To determine $r_{500}$ for each cluster, we adopt a fiducial value of the integrated gas mass fraction at this radius, $\Mgas(<r_{500})/M(<r_{500}) = 0.125$, based on the $\Mgas$--$M$ relation from \pfive{}. Statistical uncertainties on the gas mass profiles, and measurement correlations among the various observables, are propagated through this procedure to our final results. This includes the correlation induced directly by the determination of and uncertainty in $r_{500}$, which determines the measurement aperture for each quantity.

\begin{figure*}
  \centering
  \includegraphics[scale=\figscale]{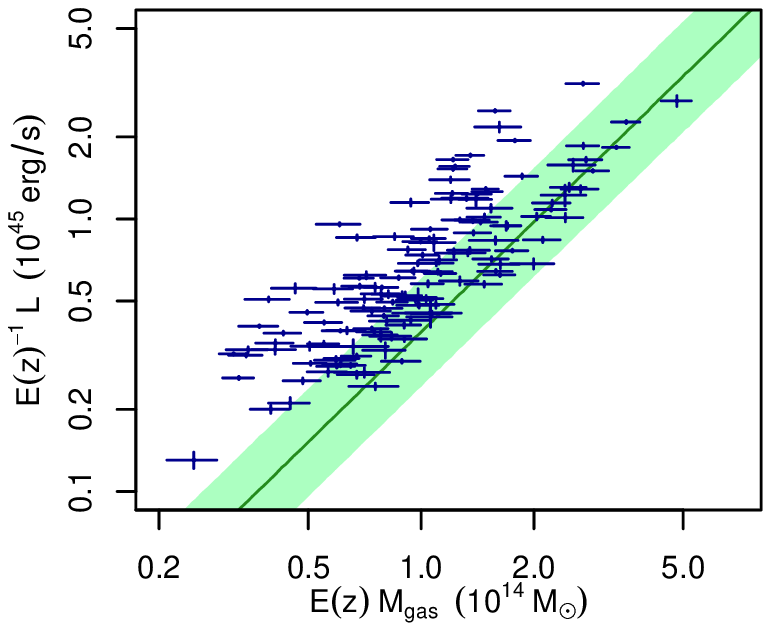}
  \hspace{1cm}
  \includegraphics[scale=\figscale]{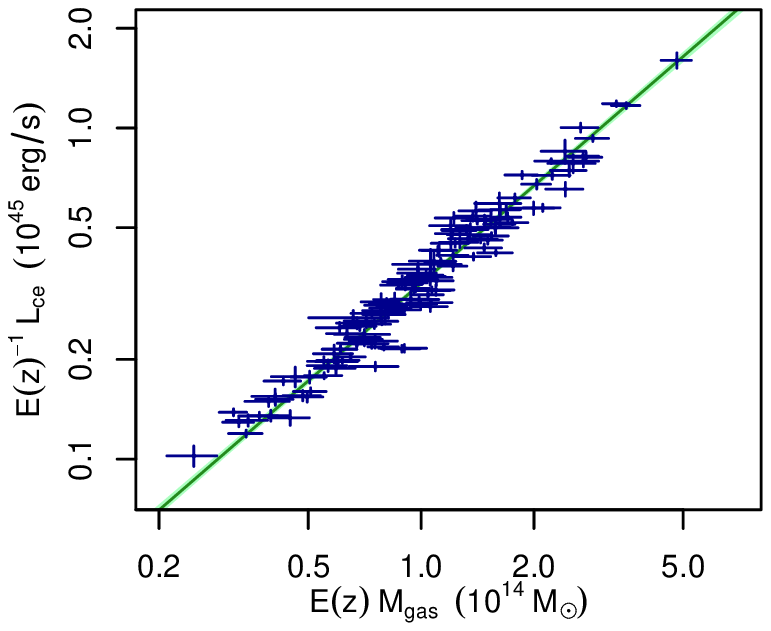}
  \caption{
    Scaling relations of $L$ (left, reproduced from \pfive) and $\Lce$ (right) with $\Mgas$. Green lines and shading show the best-fitting scaling relations and their 68.3 per cent probability predictive intervals, including intrinsic scatter. The offset between the best fit and the data in the left panel is due to selection effects (see \pfive).
  }
  \label{fig:scaling-mgas}
\end{figure*}

\section{Scaling Relations of Center-excised Luminosity} \label{sec:scaling}

Our data allow us two routes to constraining the $\Lce$--mass relation. On one hand, we have both \Lce{} and $\Mgas$ measurements for $>100$ clusters. \pfive{} showed that \Mgas{} and $M$ (calibrated from weak lensing) are tightly correlated, with a power-law slope very near unity, $M\propto\Mgas^{1.007\pm0.012}$, for the clusters in this sample. Simulations predict the intrinsic scatter in the $\Mgas$--$M$ relation to be small, $\ltsim10$ per cent \citep{Battaglia1209.4082, Planelles1209.5058, Barnes1607.04569}, a feature which has been explicitly confirmed for massive and dynamically relaxed clusters, where precise total masses can also be derived from X-ray data \citep{Mantz1402.6212, Mantz1509.01322}. We can, therefore, use the abundant gas mass measurements to fit an $\Lce$--$\Mgas$ relation (accounting for the correlation in measurement uncertainties), and then straightforwardly translate these constraints to an estimate of the $\Lce$--$M$ relation. One non-trivial aspect of this approach is the indication, from relaxed clusters, of a strong correlation in the \emph{intrinsic} scatters of $\Lce$ and $\Mgas$ at fixed mass \citep{Mantz1509.01322}; this is a direct consequence of the near equivalence of $\Lce$ and $\Mgas$ for self-similar clusters. While this will not affect the slope of the inferred $\Lce$--$M$ relation, it does mean that any measured intrinsic scatter in $\Lce|\Mgas$ may underestimate the intrinsic scatter in $\Lce|M$.

On the other hand, we have, for a smaller number of clusters, measurements of both the total mass from weak lensing and $\Lce$. The statistical uncertainties on individual weak lensing mass estimates are relatively large compared to those of \Mgas{}, as is the expected intrinsic scatter between lensing-derived mass and true mass \citep{Becker1011.1681}. Nevertheless, careful weak lensing mass estimates represent the current gold standard of absolute cluster mass calibration (i.e., accuracy in the average). Our approach in this section is, therefore, to fit the $\Lce$--$\Mgas$ scaling relation, as described above, and then to demonstrate the consistency of this relation with the weak lensing mass estimates.
 
\begin{figure}
  \centering
  \includegraphics[scale=\figscale]{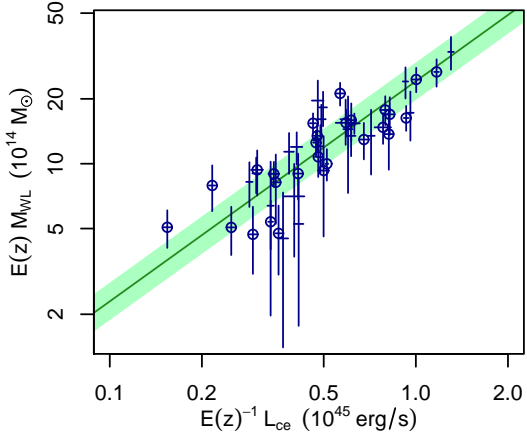}
  \caption{
    Scaling of $\Lce$ with total mass from weak lensing. Green lines and shading show the best-fitting scaling relation and their 68.3 per cent predictive intervals (including the intrinsic scatter in lensing mass), where the scaling relation is that of Figure~\ref{fig:scaling-mgas} (right) with $\Mgas$ transformed to total mass using the best-fitting $\Mgas$--$M$ relation from \pfive. All clusters with lensing masses from \citet{Applegate1208.0605}, including some that are absent from the sample used to fit the scaling relation, are shown; those that are used in the fit are circled.
  }
  \label{fig:scaling-mwl}
\end{figure}

The expected self-similar scaling of $\Lce$, following \citet{Kaiser1986MNRAS.222..323K}, has the form
\begin{equation} \label{eq:full-scaling}
  \Lce \propto E(z)^{2+\alpha\,\beta_{t}} \, M^{1+\alpha\,\beta_{t}} \propto E(z)^{2+\alpha\,\beta_{t}} \, \Mgas^{1+\alpha\,\beta_{t}},
\end{equation}
where $E(z)=H(z)/H_0$. Here $\beta_{t}$ is the power-law slope of the temperature--mass relation (from \pfive, $\beta_t=0.62\pm0.04$, consistent with the self-similar value of $2/3$), and $\alpha\approx-0.13$ is the intrinsic temperature dependence of bremsstrahlung emissivity, accurate for temperatures $\gtsim3\keV$, accounting for the limited energy band in our definition of luminosity (rest-frame 0.1--2.4\,keV). Our baseline expectation can thus be written in a simpler form,
\begin{equation} \label{eq:simple-scaling}
E(z)^{-1}\Lce \propto \left[E(z)\,M\right]^\gamma,
\end{equation}
with $\gamma=1+\alpha\,\beta_t\approx 0.92$ for $\beta_t=2/3$. The center-excised luminosity, gas mass and lensing mass data are shown in terms of this scaling in Figures~\ref{fig:scaling-mgas} and \ref{fig:scaling-mwl}, although we consider a more general dependence on redshift and mass below.

We fit scaling relations to the data using a model that includes a log-normal intrinsic scatter (i.e.\ a Gaussian distribution in $\ln\Lce$), accounts for the correlation between measurement uncertainties, and marginalizes over a log-normal distribution of the covariates \citep{Kelly0705.2774, Mantz1509.00908}. This methodology does not explicitly account for selection effects (e.g.\ Malmquist bias), but because the underlying data set is X-ray flux limited, with the majority of the intrinsic scatter in total X-ray luminosity being due to the core region that is excluded from $\Lce$ \citep{Mantz1509.01322}, we do not expect selection effects to significantly impact our results. Fitting the simpler form of the $\Lce$--$\Mgas$ relation from Equation~\ref{eq:simple-scaling}, we find
\begin{equation} \label{eq:bestfit}
  \frac{E(z)^{-1} \Lce}{10^{45}\erg\second^{-1}} = e^{(-1.081\pm0.010)}  \left[\frac{E(z) \Mgas}{10^{14}\Msun}\right]^{0.979\pm0.019}.
\end{equation}
Note that the departure of the power-law slope from the self-similar baseline is in the same sense (steeper), but smaller, than that of the center-included luminosity scaling derived from the same cluster sample (\pfive). The log-normal intrinsic scatter is consistent with zero, with a 95.4 per cent confidence upper limit of 0.04. Figure~\ref{fig:scaling-mgas} compares this best fit, and its associated 68.3 per cent confidence predictive interval, with the data, and highlights the dramatic reduction in scatter resulting from the center excision. Allowing separate power-law dependences of $\Lce$ on $E(z)$ and $\Mgas$, $\Lce \propto E(z)^{\gamma_1}\Mgas^{\gamma_2}$, we find
\begin{eqnarray}
  \frac{\Lce}{10^{45}\erg\second^{-1}} &=& e^{(-0.878\pm0.011)} \left[\frac{E(z)}{E(0.2)}\right]^{(2.03\pm0.19)}\\
  && \times \left[\frac{\Mgas}{10^{14}\Msun}\right]^{0.977\pm0.022}.\nonumber
\end{eqnarray}
This is consistent with the simpler scaling in Equation~\ref{eq:simple-scaling}; in particular, $\gamma_1-\gamma_2 = 1.06\pm0.20$.

Using the $\Mgas$--$M$ relation from \pfive, we can translate Equation~\ref{eq:bestfit} to a scaling relation between $\Lce$ and total mass:
\begin{equation} \label{eq:bestfit-Mtot}
  \frac{E(z)^{-1} \Lce}{10^{45}\erg\second^{-1}} = e^{(-0.86\pm0.04)}  \left[\frac{E(z) M_{500}}{10^{15}\Msun}\right]^{0.99\pm0.02}.
\end{equation}
Figure~\ref{fig:scaling-mwl} compares this relation with X-ray measurements of \Lce{} and weak lensing mass estimates; note that the predictive interval shown accounts for a log-normal intrinsic scatter of 0.17 in lensing masses (\pfive). To maximize the number of clusters in the comparison, we include those which have mass estimates from \citet{Applegate1208.0605}, even if they are not in the X-ray flux limited sample used to fit the $\Lce$--$\Mgas$ relation. Whether or not one makes this distinction, the scaling relation in Equation~\ref{eq:bestfit-Mtot} is in excellent agreement with the measured weak lensing masses.

While we can directly fit for the intrinsic scatter in $\Lce$ at fixed $\Mgas$, inferring the scatter in $\Lce$ at fixed total mass is more complex. For massive, relaxed clusters where total masses can be estimated from hydrostatic equilibrium, \citet{Mantz1509.01322} found a strong correlation in the joint scatters of $\Lce$ and $\Mgas$ at fixed mass (assuming a bivariate log-normal form of the scatter), with $\rho=0.88^{+0.06}_{-0.16}$; the same study found an $\Lce$--$M$ power-law slope of $1.02\pm0.09$, in excellent agreement with our current results. If the strongly correlated scatter holds for the entire cluster population, our measured scatter in $\Lce|\Mgas$ will be smaller than the scatter in $\Lce|M$. Indeed, the same study of relaxed clusters found a marginal log-normal intrinsic scatter in $\Lce|M$ of $0.17\pm0.05$. If we fit for the intrinsic scatter about the best-fitting $\Lce$--$M$ relation using the weak lensing data (approximating the statistical uncertainties in the lensing masses as log-normal) the resulting scatter estimate is $0.15 \pm 0.04$. Both of these scatter estimates contain contributions from other sources: the former from the unknown scatter in hydrostatic mass estimates for relaxed clusters, and the latter from scatter in weak lensing mass estimates due to correlated large scale structure (estimated to be $0.17\pm0.06$; \pfive, see also \citealt{Becker1011.1681}). Taken together, these lines of evidence cannot straightforwardly put a direct constraint on the intrinsic scatter in $\Lce|M$, but support a value in the range 0.1--0.2, with the lower bound given approximately by the scatter in $\Mgas|M$ and the upper bound by the estimates from hydrostatic and weak-lensing masses.

Note that the simple power-law behavior expected for \Lce{} (Equation~\ref{eq:full-scaling}) depends on the cluster temperature being $\gtsim3$\,keV. At cooler temperatures (lower masses), as the luminosity in soft energy band progressively approaches the bolometric luminosity, we would expect a change in slope of the \Lce{}--$M$ relation, although not necessarily an increase in scatter. Use of the above scaling relations in this regime, which is in any case an extrapolation beyond the data set employed here, is therefore discouraged. At even lower temperatures of $\ltsim2$\,keV, we might also expect the scaling relations to become more complex due to the appearance of strong emission lines in the soft band, in addition to the relatively greater role of complex feedback processes in these less massive systems. Given the challenge of obtaining individual mass measurements at low masses, the calibration of \Lce{} (and mass proxies in general) in this regime remains an subject for future exploration. We discuss these and other practical considerations further in Section~\ref{sec:applicability}.

For completeness, we also consider luminosity measured in an annulus spanning radii of $(0.15$--$1) E(z)^{-2/3}$\,Mpc. We denote this luminosity \Lm, since \cite{Maughan0703504} suggested this annulus definition as one that approximates the redshift dependence of $r_{500}$ but lacks the dependence on mass. \cite{Maughan0703504} found that using this annulus, as opposed to $(0.15$--$1)r_{500}$, had negligible effect on the inferred scatter in the center-excised luminosity--$\Yx$ relation. In the case of the $\Lm$--$\Mgas$ relation, we find a scatter of $0.106\pm0.015$. The difference between this result and the much smaller scatter obtained for the $\Lce$--$\Mgas$ annulus is consistent with our expectation of a strong correlation between $\Lce$ and $\Mgas$ at fixed mass (see above) due to the similarity of gas density profiles at the relevant radii, which the definition of \Lm{} does not take advantage of (though it does eliminate the scatter in luminosity due to cool cores). Note that this scatter constraint is not at odds with the results of \cite{Maughan0703504}, given the relative sizes of the scatters in $\Mgas$ and $\Yx$ found by \pfive{}.

\section{Discussion} \label{sec:discussion}

\subsection{Using \Lce{} to Estimate Mass} \label{sec:instructions}
 
In this section, we describe a straightforward procedure for estimating $r_{500}$ and $M$ from X-ray data, given an $\Lce$--$M$ scaling relation. We do not explicitly address statistical uncertainties in the scaling relation or the measured luminosities, but instead recommend propagating these by repeating the procedure for many monte-carlo realizations of each.

The starting point for a mass estimate is an X-ray surface brightness profile, ideally in a relatively soft energy band. We use the observer-frame 0.6--2.0\,keV band, which is typical in the literature. To use Equation~\ref{eq:bestfit-Mtot}, the surface brightness must be corrected for Galactic absorption and converted to intrinsic (rest-frame) 0.1--2.4\,keV band luminosity per unit solid angle. Both of these steps are, strictly speaking, temperature-dependent spectral corrections. However, for clusters in the mass and redshift range of our sample (i.e.\ where the $\Lce$--$M$ relation has been calibrated), the temperature dependence is small enough to ignore, typically. In a more thorough treatment, one could also straightforwardly make use of an empirical temperature estimate (if available), or consistently incorporate the temperature--mass relation from \pfive{}.

\begin{figure}
  \centering
  \includegraphics[scale=\figscale]{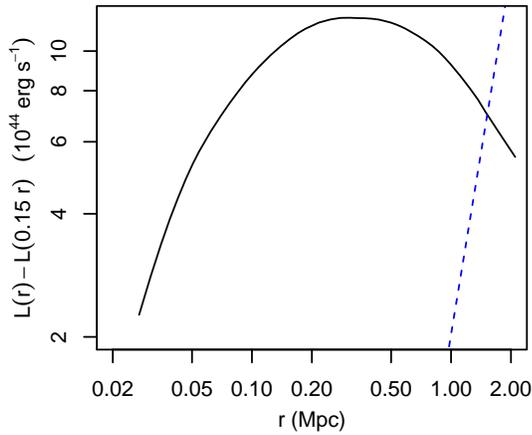}
  \caption{
    Projected aperture luminosity profile for Abell~1835. The \Lce{}-based $r_{500}$ estimate is defined by the intersection of the profile with the dashed, blue line (see Equation~\ref{eq:solveme}).
  }
  \label{fig:profile}
\end{figure}

At this point, it is straightforward to generate a projected aperture luminosity profile like the example shown in Figure~\ref{fig:profile}. As a function of radius, $r$, this profile is nothing more than the surface brightness (in luminosity units) integrated between radii of $0.15\,r$ and $r$. While the qualitative shape of the profile in Figure~\ref{fig:profile} is typical, we note that there is great variety in the aperture luminosity profiles across our sample, mostly driven by variety in the central surface brightness of the clusters. In other words, despite the self similarity of density profiles (outside cluster centers) and of the $\Lce$--$M$ relation, there is not a widely applicable scaled profile of aperture luminosity as a function of $r/r_{500}$, as there is for, e.g., $\Mgas$. In extreme cases, for very diffuse clusters with shallow central surface brightness profiles, the aperture luminosity can even be flat or increasing at $r_{500}$.

An estimate for $r_{500}$ (and hence $M$) is obtained by finding the intersection between the aperture luminosity profile and aperture luminosity predicted by the fiducial $\Lce$--$M$ scaling relation (the dashed line shown in Figure~\ref{fig:profile}). Combining Equations~\ref{eq:massdef} and \ref{eq:bestfit-Mtot}, we arrive at the implicit equation
\begin{eqnarray} \label{eq:solveme}
  \Lce &=& L(r_{500})-L(0.15\,r_{500})\\
       &=& A\,E(z) \left[\frac{2\pi E(z)\rhocr(z)}{3\E{12}\Msun} \, r_{500}^3\right]^B, \nonumber
\end{eqnarray}
where $A=(4.23\pm0.17)\E{44}\erg\second^{-1}$ and $B=0.99\pm0.02$.

The original $r_{500}$ estimates for our sample were arrived at by following an analogous procedure, using the measured gas mass profiles and the nominal $\Mgas$--$M$ scaling relation from \pfive{}. Comparing the nominal $r_{500}$ estimates from the two techniques (that is, using the mean scaling relations and measured profiles, without worrying about statistical or parameter uncertainties, or intrinsic scatter), we find a ratio of $0.99\pm0.03$ (average and standard deviation). Radius and mass estimates from $\Lce$ are thus highly consistent with those from $\Mgas$, as the tight scaling relation in Figure~\ref{fig:scaling-mgas} requires.

\subsection{Impact of the Telescope Resolution} \label{sec:psf}

From a practical standpoint, one of the essential differences between using $\Lce$ and $\Mgas$ as mass proxies is that the former is mostly determined by emission at relatively small projected radii, while the latter is most heavily influenced by the measured surface brightness at larger radii ($\sim r_{500}$) and requires relatively complex modeling (deprojection). This is an advantage for $\Lce$ in some respects, namely the exposure time required for a simple mass estimate and the ability to access the most important radii at all azimuths for relatively low-redshift, massive clusters in a single field of view. For distant or less massive clusters, however, we need to consider whether a given telescope's point spread function (PSF) can reliably permit the central portion of the cluster to be excised. If not, photons leaked from cluster centers can be expected to increase the intrinsic scatter in $\Lce$ mass estimates. There are other practical implications of PSF smoothing that we do not explicitly consider here, most notably the greater challenge of identifying and masking contaminating point sources in lower resolution images.

To estimate the impact of the PSF, we consider as an extreme case the cool core cluster hosting 3C\,186, an X-ray bright AGN in the central galaxy \citep{Siemiginowska1008.1739}. At a redshift of 1.06, this is one of the most centrally X-ray peaked massive clusters known at high redshifts. Based on the original \Chandra{} image, we simulate a suite of \Lce{} measurements of morphologically equivalent clusters with different redshifts and masses, observed with various PSF sizes, and with various fractions of $r_{500}$ excised. In general, the PSF was assumed to be Gaussian in shape; however, we verified that a Gaussian+beta model approximation to the real XMM PSF produces very similar results in this case to a Gaussian with the same half energy width (HEW). For centrally peaked sources such as 3C\,186, we find, intuitively, that the fractional contamination of the center-excised luminosity depends on the ratio of the PSF width to the size of the excised region, and not on the cluster redshift or mass explicitly. For this test case, we find that the ratio $\mathrm{HEW}/(x\theta_{500})$ must be $<0.35$ (0.40) to limit the contamination of the \Lce{} measurement to 5 (10) per cent, where $x\theta_{500}$ is the inner angular radius of the \Lce{} measurement annulus (i.e., $x$ is the fraction of $r_{500}$ excised). Note that these limiting values are a function of cluster morphology, and in particular are larger (less restrictive) for less extremely peaked systems.

At redshifts $1\ltsim z\ltsim2$, where the angular diameter distance changes relatively slowly with $z$, a maximum allowed level of contamination translates to a maximum PSF size, with little residual dependence on cluster redshift. If we adopt a limit of 5 per cent contamination for a 3C\,186-like cluster (small compared with the expected intrinsic scatter of $\Lce$ with mass), then our chosen excision size of $0.15\,r_{500}$ requires a HEW $\ltsim5''$. Currently, such fine resolution is  only provided by \Chandra{} ($\mathrm{HEW}<0.5''$). To meet our contamination requirement with XMM ($\mathrm{HEW}\approx15''$) over this redshift range would require us to raise the excision radius to $x=0.5$. This comes at the cost of excluding a significant fraction ($\sim75$ per cent) of the flux in the 0.15--1\,$r_{500}$ annulus for our toy model, although the larger effective area of XMM compared to \Chandra{} effectively compensates for this; for more typical (less strongly peaked) clusters, XMM would have an advantage in terms of the required observing time even using this larger excision region. In the context of future X-ray observatories, the most relevant PSF sizes are those of eROSITA ($\mathrm{HEW}\sim15''$ for pointed observations, $\sim30''$ when scanning), \Athena{} ($\mathrm{HEW}\sim5''$), and \Lynx{} ($\mathrm{HEW}<1''$).

\begin{figure}
  \centering
  \includegraphics[scale=1]{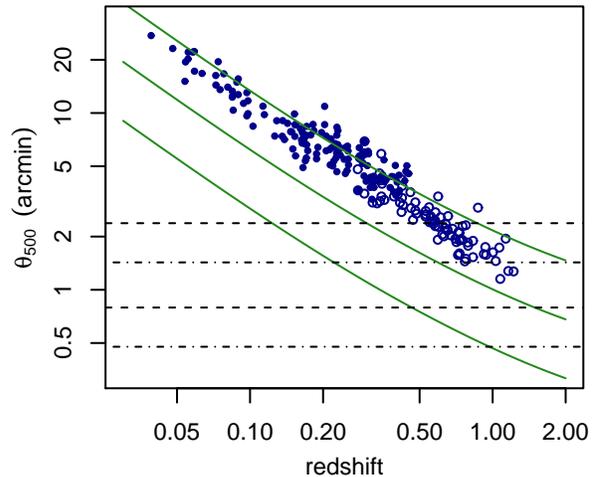}
  \caption{
    Left: values of $\theta_{500}$ are plotted against redshift for our sample (filled circles), as well as for SPT clusters extending to higher $z$ (open circles). Solid lines correspond to masses of $M_{500}=10^{13}$, $10^{14}$ and $10^{15}\Msun$, while horizontal lines show the minimum values of $\theta_{500}$ for which \Lce{} measurements are not significantly contaminated for telescopes with HEWs of $5''$ and $15''$, and using excision regions of $0.3\,r_{500}$ (dashed) or $0.5\,r_{500}$ (dot-dashed).
  }
  \label{fig:psf}
\end{figure}

Figure~\ref{fig:psf} shows the angular radii corresponding to $r_{500}$ for our sample, which comprises the most massive clusters at redshifts $<0.5$, as well as for the most massive South Pole Telescope (SPT) clusters out to higher $z$ \citep{Bleem1409.0850}. Also shown are curves corresponding to masses of $M_{500}=10^{13}$, $10^{14}$ and $10^{15}\Msun$. Dashed lines correspond to the minimum cluster $\theta_{500}$ accesible (according to the requirement above) to a telescope with a HEW of $5''$ or $15''$ (\Athena{}-like versus XMM/eROSITA-like) with an excision of $x=0.3$, while dot-dashed lines show the same for an excision of $x=0.5$. In the context of $z>1$ clusters specifically, XMM and eROSITA will be able to measure \Lce{} for massive clusters by employing an  excision radius of $x\sim0.5$. \Athena{} will have access to a larger range in mass, even with a more modest excised region. Very high resolution observatories such as \Chandra{} and \Lynx{} are essentially unrestricted. At lower redshifts, especially $z\ltsim0.5$, all the observatories considered here can access a wide range in mass.

 Given the significant practical advantages of being able to use XMM, and eventually eROSITA and \Athena{}, for \Lce{} measurements at high redshifts, we list in Table~\ref{tab:fits} scaling relation parameters corresponding to excision regions of $x=0.3$ and 0.5, in addition to 0.15. The table also includes fits where luminosity corresponds to the alternative but commonly used rest-frame band of 0.5--2.0\,keV.

\begin{table}
  \begin{center}
    \caption{
      Parameters for scaling relations of the form $E(z)^{-1} \Lce/(10^{45}\erg\second^{-1}) = e^A  \left[E(z) M_{500}/(10^{15}\Msun)\right]^\gamma$, where \Lce{} corresponds to the given rest-frame energy bands and annuli.
    }
    \label{tab:fits}
    \vspace{-1ex}
    \begin{tabular}{cccc}
      Band & Annulus & $A$ & $\gamma$ \\
      (keV) & ($r_{500}$) \\
      \hline
      0.1--2.4 & 0.15--1 & $-0.86\pm0.04$ & $0.99\pm0.02$ \\
      0.1--2.4 & 0.30--1 & $-1.49\pm0.04$ & $0.96\pm0.02$ \\
      0.1--2.4 & 0.50--1 & $-2.27\pm0.04$ & $0.96\pm0.03$ \\
      0.5--2.0 & 0.15--1 & $-1.32\pm0.04$ & $0.97\pm0.02$ \\
      0.5--2.0 & 0.30--1 & $-1.96\pm0.04$ & $0.97\pm0.02$ \\
      0.5--2.0 & 0.50--1 & $-2.75\pm0.04$ & $0.97\pm0.03$ \\
   \hline
   \end{tabular}
  \end{center}
\end{table}

\subsection{Regime of Applicability} \label{sec:applicability}

Several considerations impact the regime of redshift, temperature, and mass where we can expect the scaling relations given in Section~\ref{sec:scaling} and Table~\ref{tab:fits} to provide a good description of clusters. To begin with, the scaling relations reported here are calibrated using high quality X-ray and weak lensing observations of clusters with masses $M_{500}\gtsim 3\E{14}\Msun$ (temperatures $\gtsim4\keV$) at redshifts $<0.5$. The good agreement with relaxed clusters of similar masses out to $z\sim1.1$, for which we have reliable hydrostatic mass estimates \citep{Mantz1509.01322}, suggests that the scaling relations remain valid for massive clusters at least out to these redshifts. To extend their use to even higher redshifts and/or lower masses, additional gas mass and/or total mass measurements should be obtained in order to verify that the scaling relations remain valid.

More broadly, we can ask in what regime we theoretically expect the \Lce--$M$ relation to remain a simple power law with small intrinsic scatter. One key requirement is the approximate self-similarity of gas density profiles within the luminosity measurement aperture. This property has been repeatedly verified for massive clusters over a wide range in redshifts. Going down to the group scale, we would generally expect the sphere of influence of a central AGN to be larger relative to $r_{500}$, potentially altering the mean scaling relation and sourcing additional intrinsic scatter; however, measurements of a constant gas mass fraction between $r_{2500}$ and $r_{500}$ ($\approx0.5$--1\,$r_{500}$) by \citet{Sun0805.2320} suggest that $\Lce$ may remain well behaved for a sufficiently large excision radius.

Another requirement for the scaling relation to remain simple is that the emission in the band where \Lce{} is defined be dominated by bremsstrahlung. At temperatures $\ltsim2\keV$, line emission plays an increasingly important role. In principle, this contribution is predictable, and one could simply alter the power law expectation accordingly. However, the dependence on a potentially complex metallicity structure in poor clusters and groups likely introduces a significant scatter.

A related consideration at high redshifts and/or low masses (temperatures) is the ease of converting measured count rates in the observer-frame to intrinsic luminosity in the cluster rest frame. For $kT/(1+z)>3\keV$, the temperature dependence of K corrections in the soft X-ray bands generally used is small, making relatively shallow observations to obtain $\Lce$ a viable option for mass estimation. At sufficiently low temperatures and/or high redshifts, however, the bremsstrahlung spectral cut-off will be present within the observer-frame energy band used to measure fluxes, at which point the temperature dependence of the K corrections can no longer be neglected. (At low temperatures, line emission provides another source of temperature dependence in the K correction.) Calculation of \Lce{} in this regime thus requires observations to be deep enough to measure temperature, which provides a mass proxy in its own right. Depending on the cluster redshift and the instrumental background, the required observations could also potentially provide gas mass measurements. It remains to be seen whether \Lce{} or one of the other mass proxies, or some combination of them \citep{Ettori1111.1693, Mantz1509.01322} is most efficient in this regime.

\subsection{Exploiting New Cluster Surveys} \label{sec:surveys}

In the coming decade, new cluster surveys at X-ray, optical/NIR and mm wavelengths will vastly expand the population of clusters known. To obtain the tightest possible cosmological constraints from these new catalogs, additional data will be required to set the absolute cluster mass calibration (e.g.\ through weak lensing) and also to provide information on the evolving shape of the mass function through precise relative mass measurements. The latter task is where mass proxies such as \Lce{} can contribute, provided that their scaling relations with mass and redshift are understood. In this context, a low-scatter mass proxy that is straightforward to measure from relatively short X-ray observations is potentially invaluable. For new X-ray surveys such as eROSITA, \Lce{} can be estimated directly from survey data for many more clusters than other X-ray mass proxies such as $\Mgas$ and temperature; by the same token, follow-up observations of clusters discovered in mm-wavelength and optical surveys can be shorter if the target is to measure \Lce{} rather than the other proxies. For Sunyaev-Zel'dovich (SZ) effect surveys especially, newly discovered clusters will tend to have high temperatures, making them well suited to mass estimates based on \Lce{} (Section~\ref{sec:applicability}).

To provide a rough quantitative estimate of the impact that follow-up observations with \Chandra{}, XMM and/or future X-ray telescopes might have, we use the Fisher forecasting code of \citet{Wu0907.2690}. The fiducial survey we consider has a mass limit of $2\E{14}\Msun$ and a redshift range of $0.0<z<1.5$ and covers 2000 sq.\ deg., finding $\sim5500$ clusters. This design is approximately modelled after the union of RASS and the SPT-3G surveys, but since the forecasted improvement due to follow-up data is not very sensitive to the details, the results are more broadly applicable. A limitation that should be noted, however, is that this forecasting code assumes that the survey observable scaling relation is modeled by a power-law in mass and redshift with constant intrinsic scatter. In practice, more flexible scaling models are likely to be applied to future large surveys, in which case our simple forecasts will underestimate the value of auxiliary mass proxy information.

\begin{figure}
  \centering
  \includegraphics[scale=\figscale]{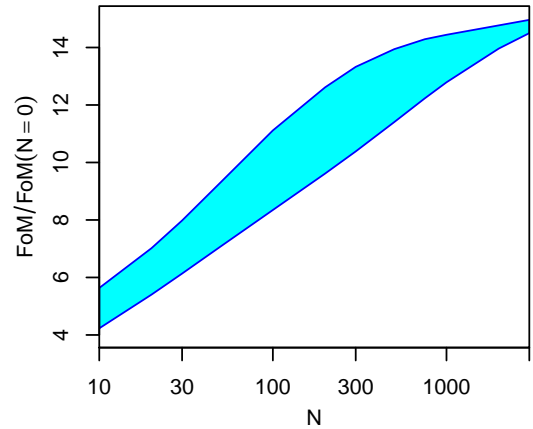}
  \caption{
    Predicted improvements in the dark energy FoM, for a fiducial survey of massive clusters extending to high redshifts, as a function of the number of clusters followed-up to provide low-scatter mass proxies. The upper edge of the shaded region corresponds to follow-up targets chosen to optimize the FoM, while the lower edge corresponds to representative follow-up of detected clusters. Both cases assume power-law scaling relations with constant scatter. In practice, the best strategy would not rely on such strong assumptions about the form of the scaling relations, but would spread follow-up observations more evenly in mass and redshift than representative sampling.
  }
  \label{fig:fish}
\end{figure}

Figure~\ref{fig:fish} shows predicted improvements in the dark energy figure of merit \citep[FoM;][]{Albrecht0609591} as a function of the number of follow-up mass proxy measurements with 15 per cent intrinsic scatter. The upper edge of the shaded region corresponds to follow-up targets that are chosen to optimize the FoM (assuming a power-law form for the scaling relation and its evolution; see \citealt{Wu0907.2690}), while the lower edge corresponds to representative follow-up of the survey. In the optimized case, roughly half of the targets are relatively low-redshift, high-mass clusters of the kind that already populate the \Chandra{} and XMM archives and will have high signal-to-noise in the eROSITA survey; hence, new observations would focus on the relatively high-redshift and low-mass clusters. Adopting a target of 100 source counts in the 0.15--1\,$r_{500}$ aperture, and accounting for the mass and redshift distribution of follow-up targets, the \Chandra{}-equivalent exposure time required is approximately 1\,Ms per 50 new cluster observations, with most individual exposures in the 10--30\,ks range.

Arguably the most beneficial follow-up strategy in terms of the potential for discovery is not one that is optimized assuming a particular model for the cosmology and scaling relations, as above, but rather one that spreads follow-up observations throughout the interesting redshift and mass range. The FoM improvement per target for such an ``evenly sampled'' program would lie between the extremes represented by the optimal and representative follow-up cases, i.e.\ order of magnitude improvements in the FoM for $\sim100$--500 mass proxy measurements in total. In this context, we note that the exposure time required per 50 clusters uniformly distributed in redshift and $\log(M)$, for $0.3<z<1.5$ and $10^{14}<M/M_\odot<6\E{14}$ (i.e., the regime not well represented in archival data), is again $\sim1$\,Ms with \Chandra{}, similar to the optimized case above. The rough scale of a follow-up program like the one outlined here is thus not too sensitive to the choice of targets, with  \Chandra{}-equivalent investments of 3--5 Ms potentially producing order of magitude improvements in the FoM with respect to no follow-up for the fiducial SPT-3G case, scaling to $\sim10$\,Ms spread over the next decade to enable more ambitious surveys such as CMB-S4.

Inevitably, relatively short observations that are optimized to measure \Lce{} will produce less auxiliary astrophysics per observation than the deeper exposures that have been the norm to date. Nevertheless, there is science beyond cosmology that such observations would enable, such as the evolution of cool cores (as identified by surface brightness), cluster morphologies, and the identification of active galactic nuclei (AGN) in and around clusters (given sufficient spatial resolution e.g.\ \citealt{Ehlert1209.2132, Ehlert1310.5711, Ehlert1407.8181}). Given that the clusters that new surveys will uncover are, naturally, X-ray fainter than those that have already been studied, it is reasonable that the first systematic forays into this new regime be focussed primarily on such basic measurements, with interesting candidates for deeper exposures being identified from these initial observations. In practice, deeper observations of a subset of clusters would be desirable in any case, to verify that scaling relations among various X-ray observables remain well behaved, as well as to exploit additional cosmological constraints from the gas mass fractions of the most dynamically relaxed clusters identified \citep{Allen1307.8152, Mantz1402.6212}. Together, these considerations suggest that, in the near term, the most efficient follow-up strategy will utilize the complementary strengths of both \Chandra{} and XMM. Relatively short \Chandra{} observations can efficiently constrain AGN populations and cluster morphologies, and measure \Lce{} (all requiring roughly equivalent exposure time). For clusters that are especially interesting in their own right, and for a subset where additional mass proxies are used to test the \Lce{} scaling relations, XMM observations (benefitting from the \Chandra{} constraints on AGN contamination) can then provide deeper imaging and more detailed spectroscopy.

In this section, we have focussed on the potential application of \Lce{} as a mass proxy for upcoming SZ surveys, but it can potentially also benefit cosmology with optically selected cluster samples, such as the Dark Energy Survey. However, because the bulk of optically selected clusters are lower in mass than the sample considered in this work, the first step in this case is to extend the calibration of the $\Lce$--$M$ relation to somewhat lower masses. In the longer term, follow-up programs analogous the one described above, but targeting the much larger sample of fainter clusters discovered by LSST, {\it Euclid} and CMB-S4, could be enabled by upcoming X-ray facilities such as {\it Athena} and {\it Lynx}.

\section{Conclusion} \label{sec:conclusion}

We present constraints on the scaling relations of the projected, center-excised X-ray luminosity, using a large sample of massive galaxy clusters with X-ray and weak lensing mass estimates. Our analysis confirms earlier indications that $\Lce$ correlates tightly with mass in the mass and redshift range probed by our data set ($M_{500}\geq3\E{14}\Msun$, $z\leq0.5$), with an intrinsic scatter of $\ltsim15$ per cent. We outline a straightforward procedure for estimating masses using this scaling relation, which requires only a cluster redshift, standard manipulations of X-ray surface brightness measurements, and the solution of an implicit equation. We comment on the spatial resolution required to take advantage of \Lce{} as a mass proxy for particularly high-redshift clusters. This is especially apt in the context of new X-ray surveys such as eROSITA, since \Lce{} can be estimated directly from survey data for many more clusters than more expensive low-scatter mass proxies such as $\Mgas$ and temperature, and for upcoming mm-wavelength surveys that will discover large numbers of high-redshift clusters. While conventionally \Lce{} has been measured in an aperture of 0.15--1\,$r_{500}$, the scatter of center-excised luminosity at fixed mass remains small with significantly more generous center excisions (e.g.\ 0.3--1 or 0.5--1\,$r_{500}$), making these observations feasible for telescopes with HEWs of $>10''$ such as XMM and eROSITA.

The relative inexpensiveness of \Lce{} compared with other X-ray mass proxies opens up the possibility of following up large numbers of clusters discovered by upcoming X-ray, optical and mm-wavelength surveys, with higher redshifts and/or lower masses than the majority of clusters in current archival data. From Fisher matrix calculations, we estimate that order of magnitude improvements in dark energy constraints from upcoming cluster catalogs are possible by investing the equivalent of $\sim$3--5\,Ms of \Chandra{} time to follow up hundreds of clusters, ideally with a comparable investment by XMM providing deeper exposures for a subset of targets. On the scale of the next decade, this represents a significant but reasonable and practical investment; indeed, our most recent cluster cosmology work, using the most massive X-ray selected clusters at $z<0.5$ (\citealt{Mantz1407.4516}; \pfive), employed almost 10\,Ms of archival \Chandra{} data, all of which has also been used in multiple astrophysical studies (and would be leveraged again in the work described here). While relatively shallow observations aimed at measuring \Lce{} do not provide very detailed thermodynamic information, they do nevertheless enable a subset of interesting and important astrophysical investigations, and could be used as a first pass to identify targets for deeper observations. Looking ahead to the late 2020s and 2030s, a similar strategy using {\it Athena} and {\it Lynx} could prove valuable for exploiting the full astrophysical and cosmological potential of cluster catalogs from LSST, {\it Euclid} and CMB-S4.

\section*{Acknowledgements}
We acknowledge support from the U.S. Department of Energy under contract number DE-AC02-76SF00515, and from the National Aeronautics and Space Administration under Grant No.\ NNX15AE12G issued through the ROSES 2014 Astrophysics Data Analysis Program.

\def \aap {A\&A}
\def \apj {ApJ}
\def \apjl {ApJ}
\def \apjs {ApJS}
\def \araa {ARA\&A}
\def \mnras {MNRAS}

\label{lastpage}
\end{document}